# μ-FlowNet: A Deep Learning Approach for Mapping Flow Fields in Irregular Microchannels Using an Attention-based U-Net Encoder-Decoder Architecture

**Ganesh Sahadeo Meshram[1], Suman Chakraborty[1]\*, Nishant Sinha[2], Partha P Chakrabarti[3]**

[1]Department of Mechanical Engineering, IIT Kharagpur, Kharagpur, 721302, India

[2]Centre of Excellence in Artificial Intelligence, IIT Kharagpur, Kharagpur, 721302, India

[3]Department of Computer Science & Engineering, IIT Kharagpur, Kharagpur, 721302, India

Corresponding Author: suman@mech.iitkgp.ac.in

**Abstract**

In the complex domain of microfluidics systems, analysing fluid flow patterns through random-shaped circular microchannels is significantly challenging task. Conventional approach of solving such problems using computational fluid dynamics often incapable due to their intensive computational requirements and high simulation times. In this study, addressing these limitations, we introduce μ-FlowNet, a deep learning framework based on the adaptable U-Net autoencoders. This model provides a data-driven approach that enhances the prediction and mapping of random-shaped circular microchannels and their corresponding fluid flow patterns.  The datasets required for the training of the model is generated by performing extensive simulations using conventional approach of computational fluid dynamics methods. The datasets are then pre-processed and accessed the required spatial and temporal features that are essential for the training. We have trained three different models based on U-Net framework namely, standard U-Net, T-Net, and U-Net with attention mechanism to compare the prediction accuracy and loss. The accuracy of the μ-FlowNet is compared using metrics of dice score and intersection over union and it shows that U-Net with attention mechanism shows the highest dice score and IoU of 0.9317 and 0.8731, respectively and shows the highest structural similarity as compared to standard U-Net and T-Net. This show that U-Net with attention mechanism serves best model to map the fluid flow pattern with random datasets on testing.

## 1. Introduction

The fluid flow through microchannels with random-shaped circular geometries has shown substantial interest due to its relevance in microfluidic applications and the advancement of lab-on-a-chip technologies [1], [2] [3]. Unlike conventional microchannels with uniform shapes, random-shaped microchannels exhibit unique flow characteristics that significantly influence fluid dynamics and transport phenomena [4], [5]. These microchannels often used in biomedical and chemical engineering fields, requires detailed study to comprehend the flow behaviour and its impact on fluid mixing efficiency, pressure drop, and heat transfer rates[6] [7]. Recent studies have highlighted the complex interplay between channel geometry and fluid flow, revealing that random roughness can enhance fluid mixing efficiency but may also increase energy consumptions due to higher pressure drops [8]. Furthermore, the effect of surface roughness and channel constriction on fluid flow patterns have been explored, showing potential applications in targeted drug delivery systems and efficient biochemical reactors [9]. These studies show significant advancements in microfluidic flow dynamics, providing insights that are useful for the design and optimization of microfluidic devices, but these studies require high experimental costs and time which then can be reduced by using deep learning models to analyse the fluid flows in such complex random-shaped circular microchannels by providing sufficient quality datasets for the training of the models [10], [11], [12].

Data-driven techniques have greatly accelerated in various fields such as chemical, biotechnology, biomedical, and aerodynamic simulations [13]. Classical machine learning methods including polynomial regression, support vector machines, and artificial neural networks, physics-informed neural networks (PINNs) were used in the early research in this fields [14], [15] [16] [17]. PINNs have transformed computational fluid dynamics by merging deep learning with the physical principles articulated by partial differential equations (PDEs) that essentially regulate fluid movement [18]. These networks not only assists that predictions conform to physical principles but also markedly decrease computing expenses relative to conventional approaches [19]. Recent innovations in PINNs have improved their ability to manage intricate situations and geometries, as shown in research by Raissi et al., which explores the efficacy of multi-fidelity modeling with datasets of differing precision [16], [20], [21]. Recent research has broadened the application of PINNs to dynamic and risk-sensitive contexts by integrating Bayesian approaches for uncertainty quantification and using the networks to forecast fluid dynamics in uneven terrains, including porous media [22]. Moreover, PINNs are progressively being incorporated with conventional computational fluid dynamics tools, establishing a novel paradigm in which machine learning not only assists but substantially enhances the efficacy of numerical simulations in fluid dynamics [23]. This synergy is illustrated in extensive reviews and applications ranging from environmental engineering to aerospace, emphasizing the versatility of PINNs in addressing diverse fluid mechanics challenges, thus signifying a notable transition towards a more cohesive approach within the scientific community [14, 15, 19].

Deep Learning (DL) methods have gained popularity recently, which has an effect on microfluidics and aerodynamics. Innovating studies by Geneva and Zabaras [24] and Ling et al. [25] created neural networks specifically designed for modeling turbulence. A hybrid deep learning model that combines turbulence simulation with representation learning was presented by Wang et al. [26] in an effort to improve prediction accuracy and capture physical quantities. Nevertheless, these techniques are frequently extremely specialized and designed for modeling turbulence. The steady flow fields were the focus of early microfluidic predictions, and models by Guo et al. [27] and Bhatnagar et al. [28] demonstrated their limitations in obeying physical laws. By employing a variety of DL networks to refine spatial and temporal data, Lee and You [29] enhanced predictions of unsteady flow. The accuracy of updated U-Net models for steady flow field prediction without the inclusion of physical constraints was investigated by Thuerey et al. [30]. In their investigation of neural networks for temporally changing turbulent flows, Srinivasan et al. [31] highlighted the potential of long short-term memory (LSTM) and multi-layer perceptron (MLP) networks, with LSTM demonstrating promising outcomes. Recent investigations have yielded significant insights into drag prediction and the interplay of surface roughness in turbulent flows [4]. Shi et al. [32] investigated data-driven regression techniques for predicting drag on rough surfaces, underscoring the superiority of kernel approaches compared to traditional linear regression, especially in their ability to model non-linear connections in turbulent flow conditions including rough surfaces. Simultaneously, Fredrich et al. [33] has highlighted the need of precisely defining surface roughness, using both experimental and computational techniques to improve the prediction accuracy of existing models related to surface texture and its influence on fluid dynamics. Finally, Yang et al. [19] concentrated on enhancing the dependability of machine learning models for drag prediction by using sophisticated methods such as active learning to improve model accuracy and generalizability. Guastoni et al. [34] examined the practicality of using transfer learning with fully convolutional network (FCN) models at different Reynolds numbers and proposed that models can be transferred between different flow conditions. Zhu et al. [35] created a deep neural network that combines physical conservation rules and attention mechanisms to improve the accuracy of flow prediction in computational fluid dynamics simulations. The network is designed to quickly and precisely predict flow patterns while ensuring that the results align with the fundamental principles of physics.

In this study, we utilize Deep Neural Network (DNN) models especially U-Net architectures to develop a mapping between random-shaped geometry and its corresponding fluid flow patterns in terms of velocity components [36]. Although there are few works on image-to-image processing tasks, such as image segmentation [37], [38], [39], [40] but previous studies primarily concentrate on medical or natural images derived from unidentified physical phenomena. These methods lack the integration of physical concepts to direct neural network training. In our work, the flow prediction challenge is distinct from ordinary image-to-image translations. This is because the flow data we use is obtained by solving corresponding governing equations and must comply with fundamental physical laws. Our work presents a new physical loss function that provides the physical coherence of predictions. This approach serves as a connection between conventional image-to-image processing tasks and the specific difficulties encountered in predicting flow fields in random-shaped microchannels. It also allows for more precise and relevant predictions by improving accuracy.

The paper is constructed as follows: In the Datasets section, we have presented the mathematical formulation of the problem and data preparation of the fluid flow in flow through random-shaped microchannels. Afterwards, we explained the model architectures and evaluation criterion in the Methods section. Subsequently, in the Results section, we analyse the performance of the U-Net models across different test cases using metrics and loss functions. Finally, we provide limitations and concluding remarks in the Conclusion section.

## 2. Methods

### 2.1 Weierstrass-Mandelbrot (W-M) function

The Weierstrass-Mandelbrot (W-M) function is particularly advantageous in fluid flow issues, notably in simulating turbulent flows and interactions with rough surfaces [41]. Turbulence, defined by its chaotic and random characteristics, poses considerable difficulties in comprehension and forecasting, primarily because of its intricate, multiscale dynamics [42]. The W-M function, with intrinsic fractal characteristics, effectively captures the self-similar structure of turbulence across various scales [43]. Furthermore, surface roughness significantly impacts fluid dynamics, affecting boundary layer formation and drag properties in both natural and artificial systems. Conventional models often fail to precisely depict the stochastic and uneven characteristics of surface roughness. The capacity of W-M function to accurately simulate rough surfaces is essential. It allows more precise predictions of fluid dynamics at the boundary layers, improving the comprehension of phenomena like drag reduction and flow separation [44]. The W-M function offers a more robust theoretical framework for understanding fluid flows across intricate surfaces and enhances the design and optimization of engineering systems where fluid dynamics are essential. The W-M function is a fractal used to represent intricate, self-similar formations. The general form of W-M function integrates complexity and scaling, distinguished by its capacity to display analogous patterns over many scales, making it especially apt for characterizing events that defy conventional Euclidean geometry [45]. The mathematical representation of the W-M function is as follows:

$$W_H(x) = \sum_{n=-\infty}^{\infty} \frac{\cos(a^n x)}{a^{nH}} \tag{1}$$

In this context, $x$ denotes the variable, $a$ signifies a real integer beyond 1, $H$ represents the Hurst exponent, often within the range (0,1), indicating the extent of roughness or smoothness of the fractal, and $n$ indicates the summation index ranging from negative infinity to positive infinity.

**2.2 Data preparation**

In order to examine the impact of geometric shape on the random surface, we have taken into consideration a straightforward steady-state incompressible fluid flow issue for pipe flow with randomly generated cross-sections using the parametric Weierstrass-Mandelbrot (W-M) function [45] in which the radius of the circle is shown as a random function. The following represents the parametric equations of the radius with random W-M function:

$$x = R(\phi)\cos t \tag{2}$$

$$y = R(\phi)\sin t \tag{3}$$

Where $R(\phi)$ is represented as

$$R(\phi) = A^{(D-1)} \sum_{n=0}^{\infty} \gamma^{-(2-D)n} \cos(2\pi\gamma^n \phi) \tag{4}$$

Here, $D$ is the fractal dimension (1<$D$<2), A is the scaling constant, $\gamma$ is the spectral exponent, $n$ is the spatial frequency resolution, and $\phi$ is the period.

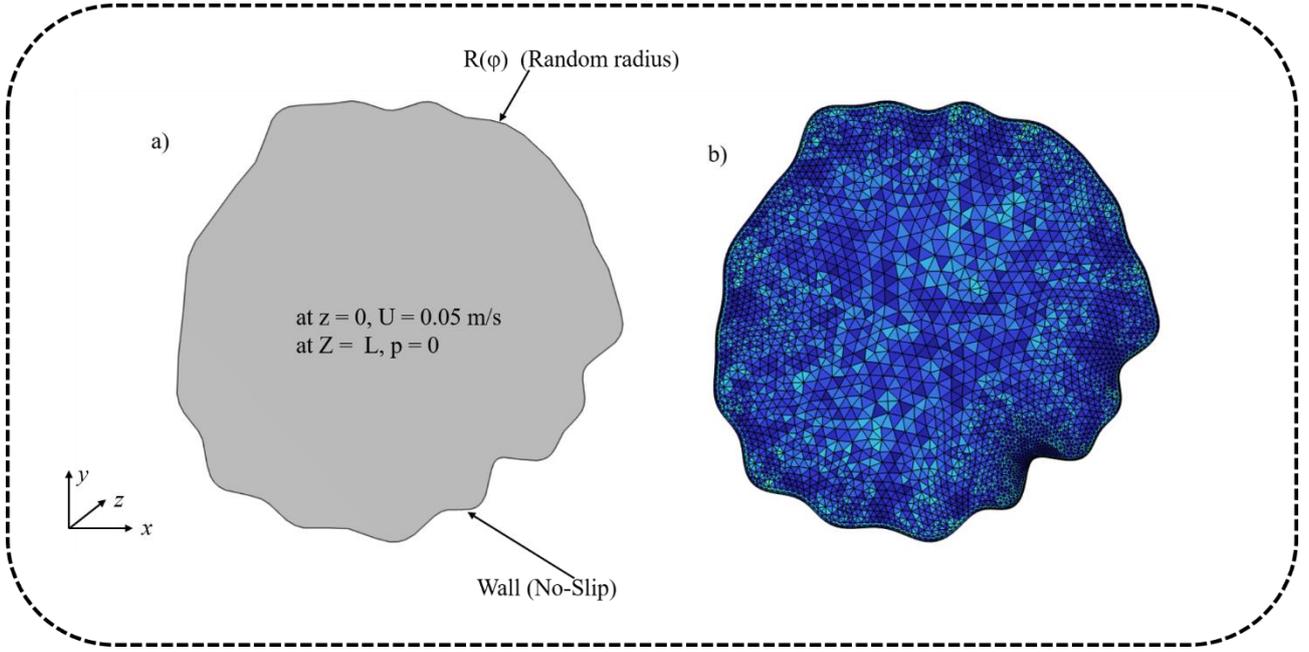

*Figure 1 a) Problem definition and b) corresponding mesh. Here U, is inlet velocity, and p is pressure outlet.*

Using fractal theory and a parametric equation, we have run 1334 simulations for various random-shaped geometries with closed cross-sections by changing the spectral exponent and spatial frequency resolution values. By fitting a spline function through randomly produced surfaces, we were able to construct an irregular wall (see Fig. 1). The mean velocity of 0.05 m/s is applied at $z = 0$ and $U$, $V_{wall} = 0$, i.e. no-slip and no penetration boundary conditions, on the walls. At $z = L$, or zero, the pressure outlet is fixed at the exit. The Navier-Stokes equations, which are the equations for the conservation of mass and momentum for a steady, 2-D, incompressible, laminar flow with no energy generation and negligible body forces, were solved to get the velocity fields [3].

$$\frac{\partial u}{\partial x} + \frac{\partial v}{\partial y} = 0 \tag{5}$$

$$\rho\left(u\frac{\partial u}{\partial x} + v\frac{\partial u}{\partial y}\right) = -\frac{\partial p}{\partial x} + \mu\left(\frac{\partial^2 u}{\partial x^2} + \frac{\partial^2 u}{\partial y^2}\right) \tag{6}$$

$$\rho\left(u\frac{\partial v}{\partial x} + v\frac{\partial v}{\partial y}\right) = -\frac{\partial p}{\partial y} + \mu\left(\frac{\partial^2 v}{\partial x^2} + \frac{\partial^2 v}{\partial y^2}\right) \tag{7}$$

Using the Parametric Curve option in the Geometry section and setting the parameters of W-M function in the Global Definitions section of COMSOL Multiphysics 6.0, the roughness profile is qualitatively represented [46]. The finite element technique-based program is used to simulate and solve the governing equations along with the boundary conditions. The domain is then discretized into mesh elements, where the solutions are approximated using polynomial shape functions, maintaining accuracy and computational cost through adaptive mesh refinement. Based on the Galerkin finite element approach, this software offers a flexible solver for multiscale and multiphase flows with nonlinear partial differential equations i.e. N-S equation [47].

## 3. Deep learning approach using U-Net

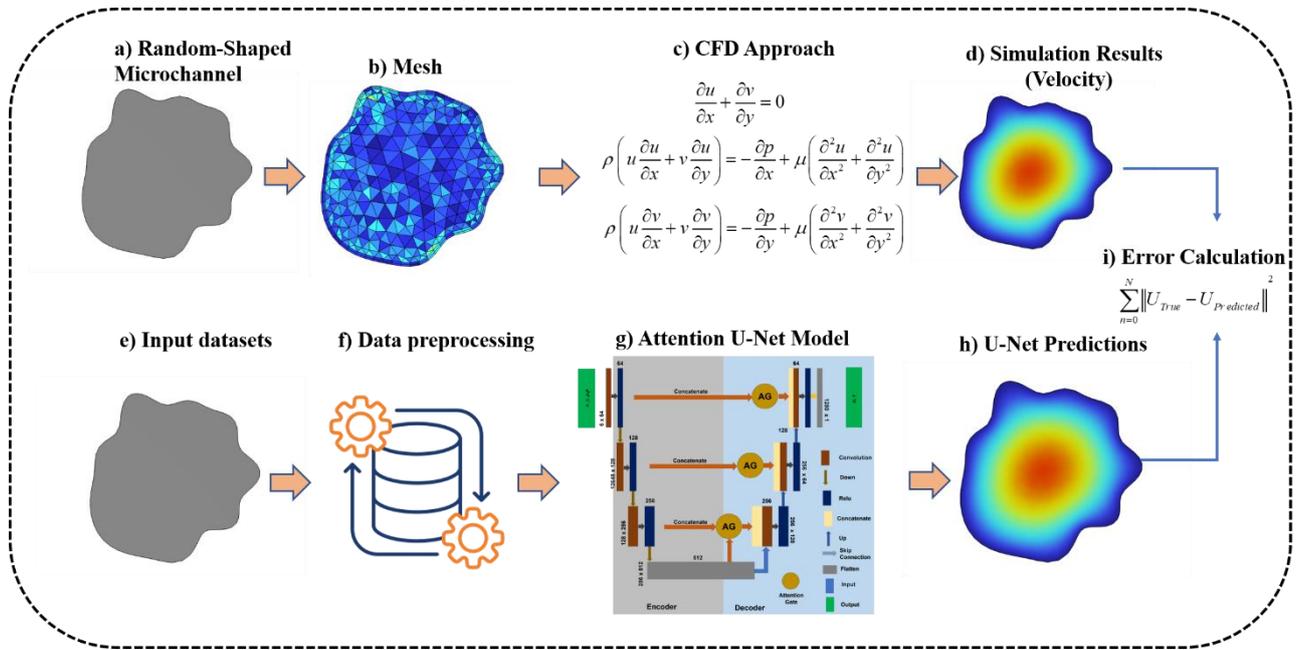

Figure 2 Comprehensive methodology for predicting fluid flow in random-shaped microchannels: Integration of computational fluid dynamics (CFD) simulations and U-Net deep learning models. (a) Geometry of a microchannel with an irregular shape used as input for the simulation or prediction task, (b) Finite element meshing of the microchannel used for solving the flow physics using numerical simulations, (c) Unstructured mesh generation, (d) Numerically obtained velocity field from the meshed domain, used as ground truth for training and comparison, (e) Geometry used as input for a data-driven model pipeline instead of simulation, (f) data pre-processing and feature

*selection, (g) Attention U-Net architecture used for predicting the velocity field directly from input geometry, integrating encoder-decoder layers, skip connections, and attention gates (AG) for enhanced spatial feature learning, (h) Predicted velocity field using the trained Attention U-Net model showing excellent agreement with the simulation results, (i) Error distribution in the Attention U-Net prediction indicating high accuracy and generalization of the model.*

The methodology presented in the *Figure. 2* illustrates a hybrid computational approach for predicting fluid flow in random-shaped microchannels, integrating traditional Computational Fluid Dynamics (CFD) with deep learning techniques. In the conventional CFD workflow, the process begins with a random-shaped microchannel geometry that undergoes discretization into a triangular mesh structure suitable for numerical analysis. This mesh serves as the spatial domain for solving the governing Navier-Stokes equations, specifically the continuity equation ($\partial u/\partial x + \partial v/\partial y = 0$) ensuring mass conservation, and the momentum equations that account for fluid inertia, pressure gradients, and viscous forces in both x and y directions. The CFD simulation yields comprehensive flow variable distributions (typically velocity fields) visualized through color maps where red regions indicate high-intensity values and blue regions represent low-intensity values. These simulation results serve as the ground truth for subsequent error calculation and validation processes.

Parallel to the traditional CFD approach, the methodology implements a U-Net deep learning architecture to predict the same flow variables with significantly reduced computational expense. The U-Net framework consists of three primary components: an encoder that progressively extracts and compresses features from the input microchannel geometry, a latent space representation that encodes essential geometric characteristics, and a decoder that reconstructs spatial information to generate predictions of flow variables. This deep learning approach bypasses the computationally intensive numerical solving of differential equations by learning the complex mapping between geometry and resulting flow patterns from training data. The accuracy of the U-Net predictions is quantitatively assessed through an error calculation formula that compares the deep learning predictions against the

CFD-generated ground truth across N samples, enabling rigorous validation while maintaining the computational efficiency advantages of neural network inference over traditional simulation methods.

**3.1 U-Net architectures**

*Figure 3* convolutional neural network architecture that is frequently used for image segmentation tasks is the U-Net model. It is made up of a symmetric expanding path that allows for exact localization and a contracting path that captures context. A sequence of down-sampling blocks, each with two convolutional layers, batch normalization, and *ReLU* activation come together to form the contracting route. *MaxPooling* is used to accomplish *DownSampling* and *ConvTranspose2D* is used for *UpSampling*.

$$\text{DownBlock}(x) = \text{MaxPool2D}(\text{ReLU}(\text{BatchNorm}(\text{Conv2D}(x)))) \tag{8}$$

The bottleneck consists of two convolutional layers with batch normalization and *ReLU* activation, which captures high-level features.

$$\text{Bottleneck}(x) = \text{ReLU}(\text{BatchNorm}(\text{Conv2D}(x))) \tag{9}$$

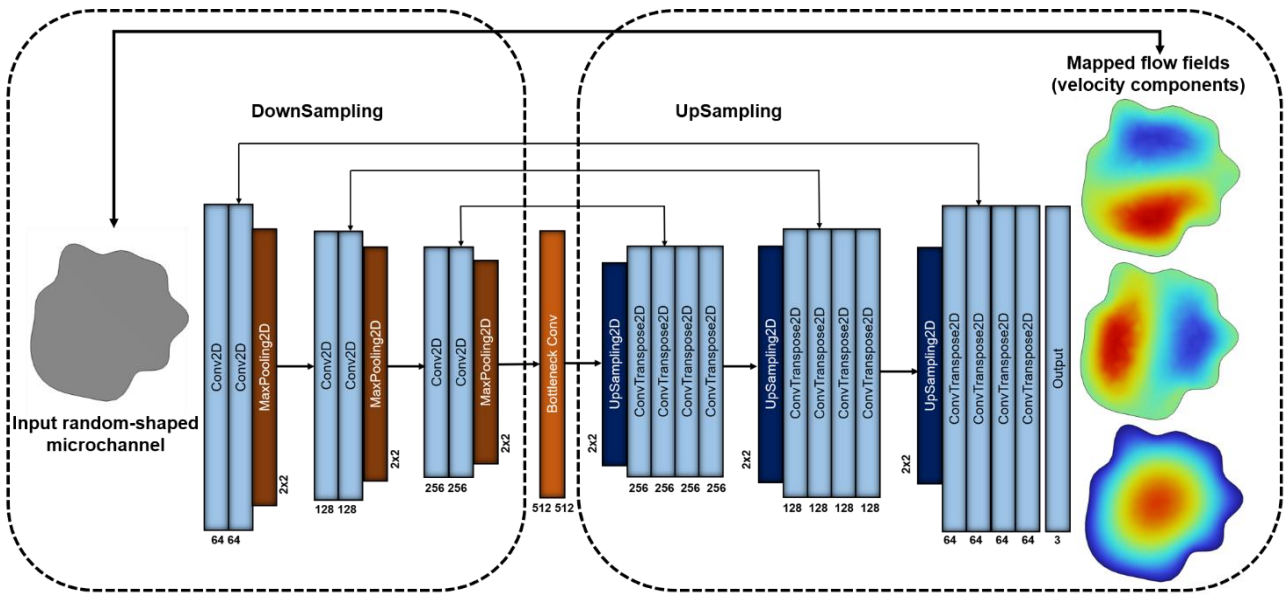

*Figure 3 Representation of U-Net model.*

The expanding path consists of up-sampling blocks, each combining features from the contracting path through skip connections as shown in *Figure 4*. Up-sampling is performed either using convolution transpose or bilinear interpolation.

$$\text{UpBlock}(x, y) = \text{ConvTranspose2D}(\text{ReLU}(\text{BatchNorm}(\text{Conv2D}(x, y)))), \text{Concatenation} \quad (10)$$

The final convolutional layer outputs the predicted flow field, with the number of output channels corresponding to the number of flow field variables.

$$\text{Output}(x) = \text{Conv2D}(x) \quad (11)$$

*Table 1 Steps involved in the U-Net architecture*

| Operations | Representation |
| --- | --- |
| Conv2d(x) | 2D Convolutional operation |
| BatchNorm(x) | Batch normalization operation |
| ReLU(x) | Rectified Linear Unit activation function |
| MaxPool2d (x) | 2D max-pooling operation |
| ConvTranspose2d(x,y) | 2D transposed convolutional operation with stride 2 |
| Concatenation | Concatenation of feature maps |

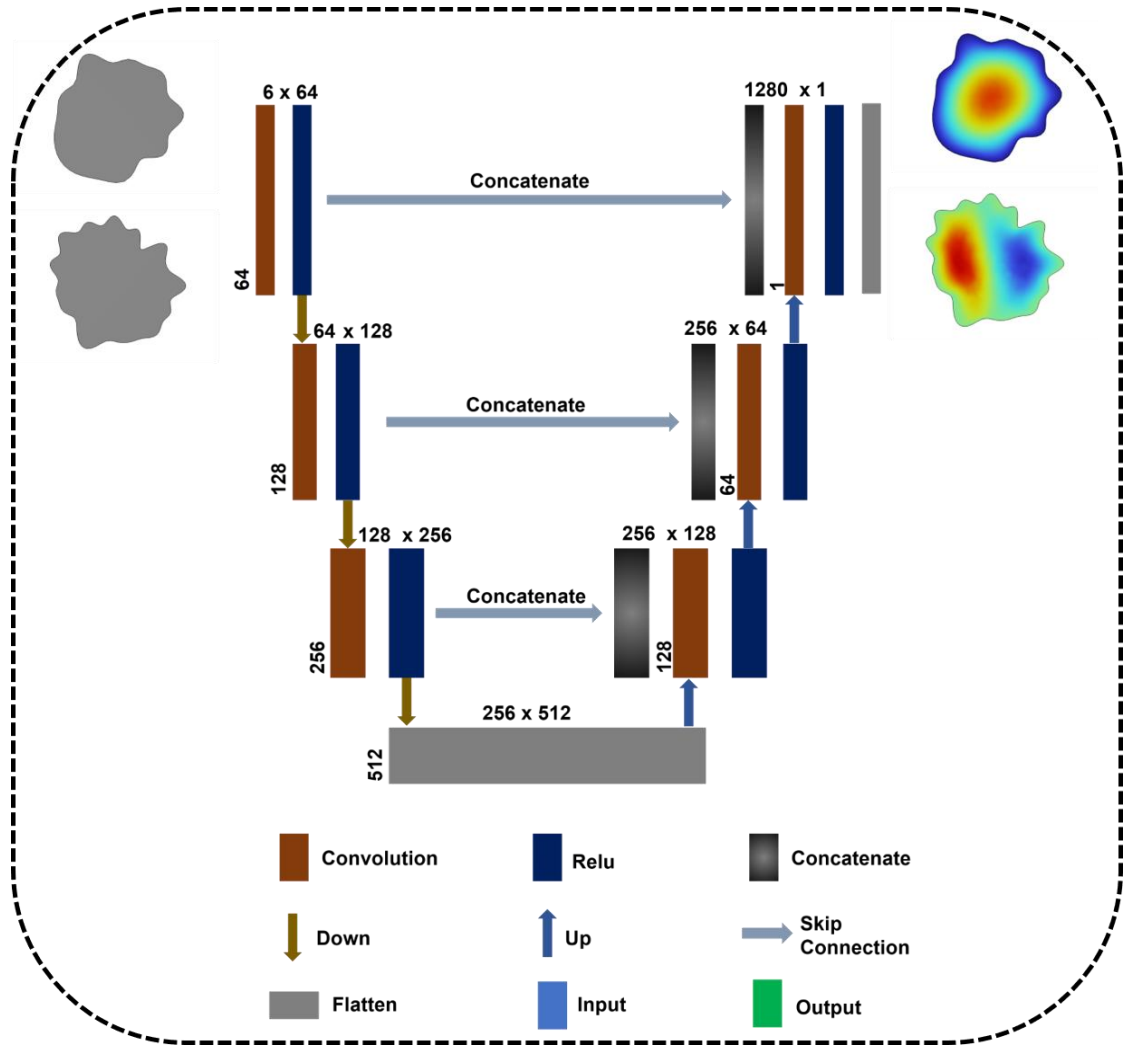

*Figure 4 Architecture of the U-Net model for the mapping of fluid flow. This encoder-decoder network uses convolutional layers and max-pooling in the contracting path to extract multiscale features, followed by upsampling layers in the expansive path to restore spatial resolution. Skip connections transfer high-resolution features from the encoder to the decoder, enhancing localization.*

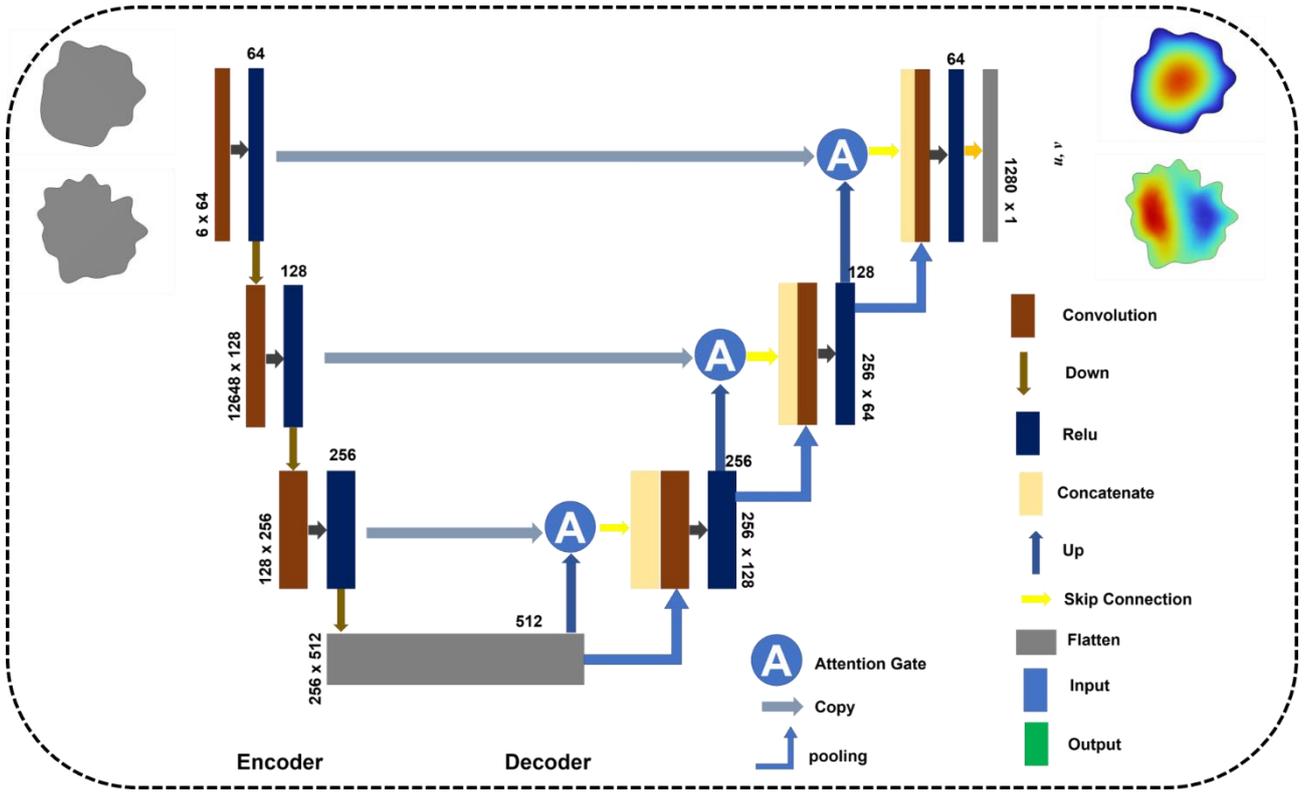

*Figure 5 Architecture of the attention-based U-Net model for the mapping of random-shaped microchannels and their corresponding fluid flow components. The network extends the standard U-Net by incorporating Attention Gates (AGs), which enhance feature selection in skip connections by suppressing irrelevant activations and highlighting important spatial features. The encoder-decoder structure captures multiscale spatial features through convolution, pooling, and upsampling operations, while AGs use contextual gating signals from the decoder to refine the feature fusion process.*

### 3.2 U-Net with Attention

*Figure 5* illustrates the overall architecture of μ-FlowNet model originally developed by Donglin et al. [36]. Our model takes input as a matrix that describes a 2D geometry domain (of size $128 \times 256$ in this work) of the target object. To generate the input matrix, we first divide the fluid domains into Cartesian grids from which we map the input fluid domain image to a matrix of 0 and 1. The model predicts the steady flows around the object given an artificial image that represents flow fields and boundaries. The model produces two matrices of size $128 \times 256$ with numerical values, where a matrix represents the velocity field for the *x*- or *y*-direction. At the core of μ-FlowNet is a U-Net architecture for steady flow prediction around arbitrary objects. U-Net was traditionally used in image segmentation to determine the area to which a pixel belongs. In this work, we extend U-Net to predict flow quantities with physical consistency for each pixel. This is achieved by using a physical loss function to constrain the training process. Unlike classical U-Net for image processing that uses a pooling layer for downsampling, we adopt a transposed convolutional kernel with a stride of 2 for downsampling. This allows the network to adjust the filter weights used for each pixel to enable a more accurate prediction at the pixel level.

µ-FlowNet has a typical U-shaped structure and comprises mainly two parts, the left part includes seven encoder blocks and the right part has seven decoder blocks. Each encoder block is followed by a convolutional layer, an activation unit, and a batch normalization layer. The convolutional kernels have a size of $4 \times 4$, except for the one in the last encoder block because the size of its input feature map is only $1 \times 2$. For each decoder block, we set up an upsampling layer followed by an activation unit. The encoder and the decoder are connected through a skip architecture, which concatenates all down-sampled feature maps from the encoder blocks to the corresponding maps in decoder blocks and doubles the number of channels. We also extend the canonical U-Net architecture by introducing attention modules (AMs), including a channel attention module (CAM) at the bottom and six spatial attention modules (SAMs) at all other skip connections. These AMs help the skip architecture integrate the fine-grained and coarse-grained information more effectively. As a departure from all prior work on DL-based CFD approximation, we introduce attention mechanisms to our learning framework. This is motivated by the observation that some RoIs in fluid flows often contain more important and complicated information than others as flow quantities change rapidly. To achieve accurate predictions in these areas, we introduce the self-attention mechanism to direct the networks to focus on RoI areas. Specifically, µ-FlowNet adopts two lightweight AMs, CAM and SAM, which are extended from Woo et al. [48]. CAM and SAM can extract the discriminative features from the channel and the spatial domains respectively to facilitate µ-FlowNet by learning which information (e.g., boundary information) to emphasize. The following equations show how these two AMs work

$$Fc = Mc(F) \otimes F \qquad (12)$$

$$Fs = Ms(F) \otimes F \qquad (13)$$

$$Mc(F) = \sigma(MLP(GAP(F)) + MLP(GMP(F))) \qquad (14)$$

$$Ms(F) = \sigma(Conv(GAPc(F) \oplus GMPc(F))) \qquad (15)$$

where $\otimes$ and $\oplus$ denote element-wise vector multiplication and channel-wise concatenation, respectively. $F \in R^{C \times H \times W}$ indicates the input feature map, while $Mc \in R^{C \times 1 \times 1}$ and $Ms \in R^{1 \times H \times W}$ represent the CAM and SAM, respectively. The intermediate results Mc(F) and Ms(F) need $\otimes$ with F itself, matching the dimension of the original input and obtaining the outputs Fc and Fs. Equations 12 to 15 show the details of operations in CAM and SAM. CAM first creates a global average pooling (GAP) and a global max pooling (GMP) along the spatial axis on the input feature map, producing a channel vector. The vector is then sent to an MLP with one hidden layer to estimate attention across channels. SAM also includes global pooling operations, but they are performed along the channel axis, GAPc and GMPc. The results from GAPc and GMPc are concatenated and sent to a convolution operation to generate a spatial attention map with one channel. Both CAM and SAM are followed by the sigmoid function σ for normalization.

### 3.3 T-Net

The network is fully convolutional with 14 layers, and consists of a series of convolutional blocks. All blocks have a similar structure: activation, convolution, batch normalization and dropout. Instead of transpose convolutions with strides, we use a linear *upsampling* "*up* ()" followed by a convolution on the upsampled content. In addition, the kernel size is reduced by one in the decoder part to ensure uneven kernel sizes, i.e., convolutions with symmetric kernels. Convolutional blocks *C* below are parametrized by an output channel factor *c*, kernel size *k*, stride *s*, where we use $c_X$ as short form for $c = X$. Note that the input channels for the decoder part have twice the size due to the concatenation of features from the encoder part. Batch normalization is indicated by *b* below. Activation by *ReLU* is indicated by *r*, while *l* indicates a leaky *ReLU* with a slope of 0.2. Slight dropout with a rate of 0.01 is used for all layers. Thus, e.g., for $ci = 6$ a *C* block with c8 has 512 channels. Channel wise concatenation is denoted by "*conc()*". Addendum: Note that l5 below inadvertently used a *kernel size* of 2 in our original implementation, and is listed as such here. While for symmetry with the decoder part, *k4* would be preferable here, this should not lead to substantial changes in terms of inference results. The network receives an input l0 with three channels and can be summarized as [49]:

$$l_1 \leftarrow C(l_0, c1k4s2)$$
$$l_2 \leftarrow C(l_1, c2k4s2lb)$$
$$l_3 \leftarrow C(l_2, c2k4s2lb)$$
$$l_4 \leftarrow C(l_3, c4k4s2lb)$$
$$l_5 \leftarrow C(l_4, c8k2s2lb)$$
$$l_6 \leftarrow C(l_5, c8k2s2lb)$$
$$l_7 \leftarrow C(l_6, c8k2s2l)$$
$$l_8 \leftarrow \mathrm{up}(C(l_7, c8k1s1rb))$$
$$l_9 \leftarrow \mathrm{up}(C(\mathrm{conc}(l_8, l_6), c8k1s1rb))$$
$$l_{10} \leftarrow \mathrm{up}(C(\mathrm{conc}(l_9, l_5), c8k3s1rb))$$
$$l_{11} \leftarrow \mathrm{up}(C(\mathrm{conc}(l_{10}, l_4), c4k3s1rb))$$
$$l_{12} \leftarrow \mathrm{up}(C(\mathrm{conc}(l_{11}, l_3), c2k3s1rb))$$
$$l_{13} \leftarrow \mathrm{up}(C(\mathrm{conc}(l_{12}, l_2), c2k3s1rb))$$
$$l_{14} \leftarrow \mathrm{up}(C(\mathrm{conc}(l_{13}, l_1), k3s1r))$$

Here $l_{14}$ represents the output of the network, and the corresponding convolution generate 3 output channels.

## 4. Model Evaluation

### 4.1 Loss function

$$L1 = \frac{1}{2mn_x n_y} \sum_{l=1}^{m}\sum_{i=1}^{n_x}\sum_{j=1}^{n_y} \left|u_{ij}^l - \bar{u}_{ij}\right| + \left|v_{ij}^l - \bar{v}_{ij}\right| \tag{16}$$

$$L1 = \frac{1}{2mn_x n_y} \sum_{l=1}^{m}\sum_{i=1}^{n_x}\sum_{j=1}^{n_y} \left|V_{ij}^l - \bar{V}_{ij}\right| \tag{17}$$

where m is the batch size, *l* denotes a certain sample, and $n_x$ and $n_y$ are the numbers of cells (pixels) along the *x*- and *y*-direction, respectively. *u* and *v* are the flow components of the *x*- and *y*-direction, respectively, and *u* and *v* stand for the predicted flow components and *V* stands for magnitude velocity.

**4.2 Evaluation Metrics**

We use the mean relative error (MRE), Dice Coefficient and IOU to evaluate the overall prediction accuracy for all flow fields. Equation (17) calculates the Mean Relative Error (MRE) as the average discrepancy between the true magnitude of the velocity field $V_{ij}$ and the predicted magnitude velocity field $\hat{V}_{ij}$ across all grid points (i, j).

$$\text{MRE} = \frac{1}{N} \sum_{l=1}^{N} \frac{\sum_{i=1}^{nx}\sum_{j=1}^{ny}\left|V_{ij} - \hat{V}_{ij}\right|}{\sum_{i=1}^{nx}\sum_{j=1}^{ny}\left|V_{ij}\right|}. \tag{18}$$

Dice Coefficient (or Dice Similarity Coefficient) is a metric used to quantify the overlap between two sets, in our case, predicted flow fields and ground truth flow fields. In theory it is calculated as follows:

$$\text{Dice} = \frac{2\sum_{i=1}^{nx}\sum_{j=1}^{ny}\left|V_{ij} \cdot \hat{V}_{ij}\right| + \varepsilon}{\sum_{i=1}^{nx}\sum_{j=1}^{ny}\left|V_{ij}\right| + \varepsilon} \tag{19}$$

where $V_{ij}$ and $\hat{V}_{ij}$ represent the sets of pixels in the predicted and actual flow fields, respectively. It calculates DICE between the true ($U_{ij}, V_{ij}$) velocity in the x-direction and y-direction respectively, and predicted components

Intersection over Union (IoU) measures the overlap between the predicted and ground truth velocity fields and is calculated as follows:

$$\text{IOU} = \frac{\sum_{i=1}^{nx}\sum_{j=1}^{ny}\left|V_{ij} \cdot \hat{V}_{ij}\right|}{\sum_{i=1}^{nx}\sum_{j=1}^{ny}\left|V_{ij}\right| + \sum_{i=1}^{nx}\sum_{j=1}^{ny}\left|\hat{V}_{ij}\right| - \sum_{i=1}^{nx}\sum_{j=1}^{ny}\left|V_{ij} \cdot \hat{V}_{ij}\right|} \tag{20}$$

where $V_{ij}$ and $\hat{V}_{ij}$ represent the sets of pixels in the predicted and actual flow fields, respectively.

**4.3 Model Training**

The models were trained on an NVIDIA GeForce RTX 090 GPU with PyTorch 2.1.1. We train the model with the adaptive moment estimation (*Adam*) optimizer [50]. To converge to stable results without overfitting, the training proceeds up to 100 epochs for *U*, *v* prediction and 75 epochs for magnitude prediction. We set the initial learning rate at $4 \times 10^{-4}$ and decay it every 25 epochs by a decay factor of 0.9, the batch size is set to 16 as shown in *Table 2*. Note that transposed convolutions may cause checkerboard artefacts Odena et al. [51]. Thus, we set the kernel size divisible by the stride to avoid this drawback.

For predicting flow magnitude, the U-Net with attention mechanism model demonstrated exceptional performance across multiple metrics as shown in *Figure 6*. It not only attained the lowest training and validation losses 0.00082 and 0.00085, respectively and mean relative errors (MRE), 0.00012 and 0.00013, respectively but also showcased impressive scores in terms of Dice coefficient and Intersection over Union. Specifically, it achieved Dice coefficients of 0.92639 for training and 0.92636 for validation, along with Intersection over Union scores of 0.86289 for training and 0.86281 for validation. Remarkably, despite its substantial parameter count of 37M, the U-Net with AM model maintained a reasonable runtime of 2 hours, 25 minutes, and 20 seconds, underscoring its computational efficiency alongside its outstanding predictive capabilities.

*Table 2 Hyperparameter setting and tuning range*

| Hyper-parameter | Value |
|---|---|
| Initial learning rate | $4 \times 10^{-4}$ |
| Decay factor | 0.9 |
| Batch size | 16 |

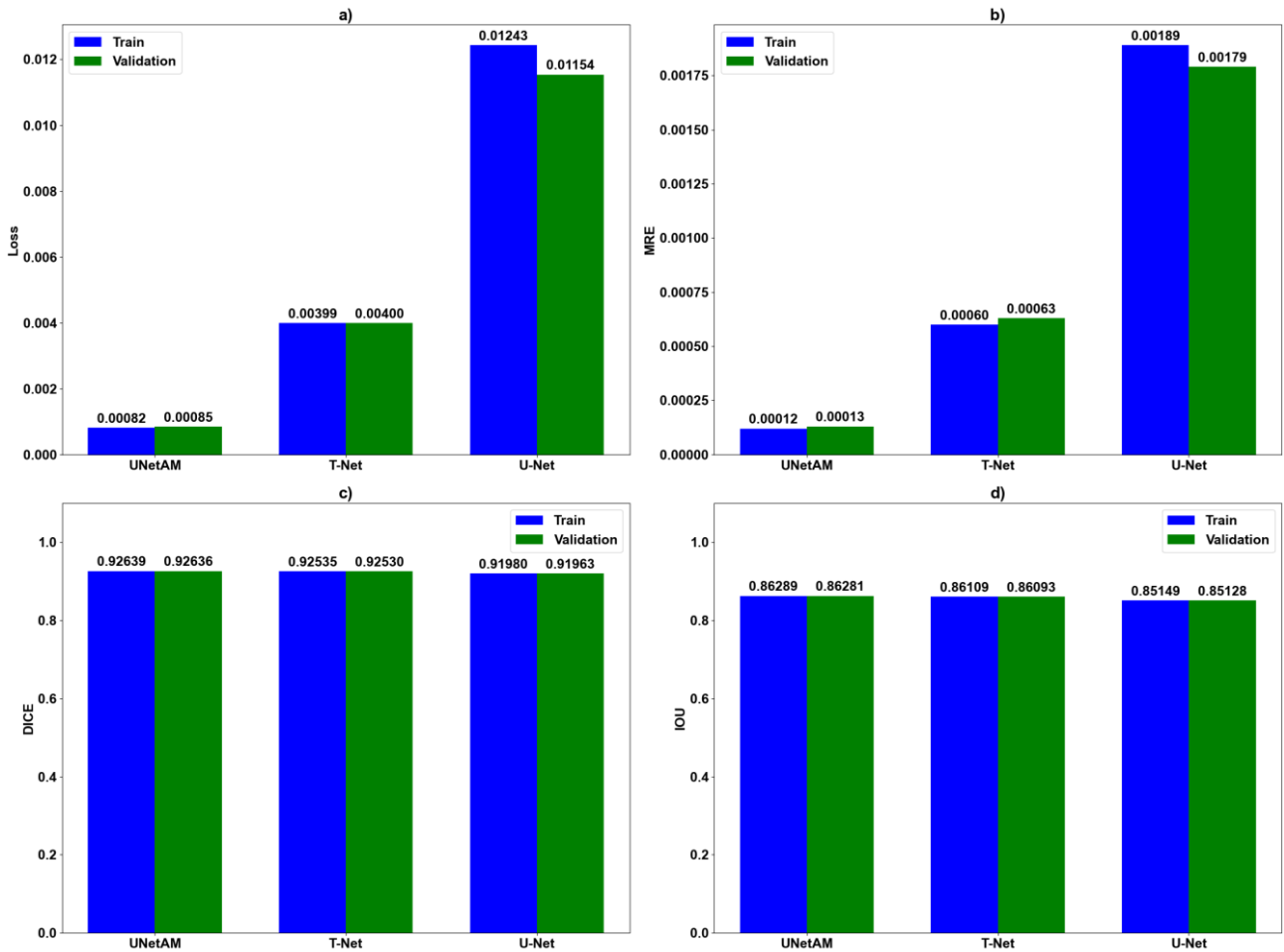

*Figure 6 Comparison of metrics of models during training and validation, a) Loss, b) mean relative error, c) dice score, and d) intersection over union.*

In the prediction of *U* and *V* components, the U-Net with AM model continued its dominance over other models i.e. standard U-Net and T-Net. Once again, it delivered the lowest training and validation losses 0.00321 and 0.00369 and MRE 0.000024 and 0.00028, highlighting its consistency and accuracy in predicting these crucial flow parameters (shown in *Figure 6*). Furthermore, its performance in terms of Dice coefficient 0.92687 for training and 0.92691 for validation and Intersection over Union 0.86371 for training and 0.86379 for validation remained highly competitive, emphasizing its robustness across various evaluation criteria. However, it is worth noting that this superior performance came at the expense of a longer runtime, total 4 hours, 43 minutes, and 9 seconds, indicating the trade-off between computational resources and predictive accuracy. The T-Net model emerged as a strong competitor, demonstrating competitive results with slightly higher losses and MRE compared to the U-Net model. Despite these minor differences, the T-Net model boasted respectable scores in terms of Dice coefficient 0.92535 for training and 0.92530 for validation and Intersection over Union 0.86109 for training and 0.86093 for validation), underlining its effectiveness in flow prediction tasks. Notably, the T-Net model exhibited a shorter runtime of 2 hours, 25 minutes, and 44 seconds, offering a more expedient computational solution without significant compromise in predictive performance.

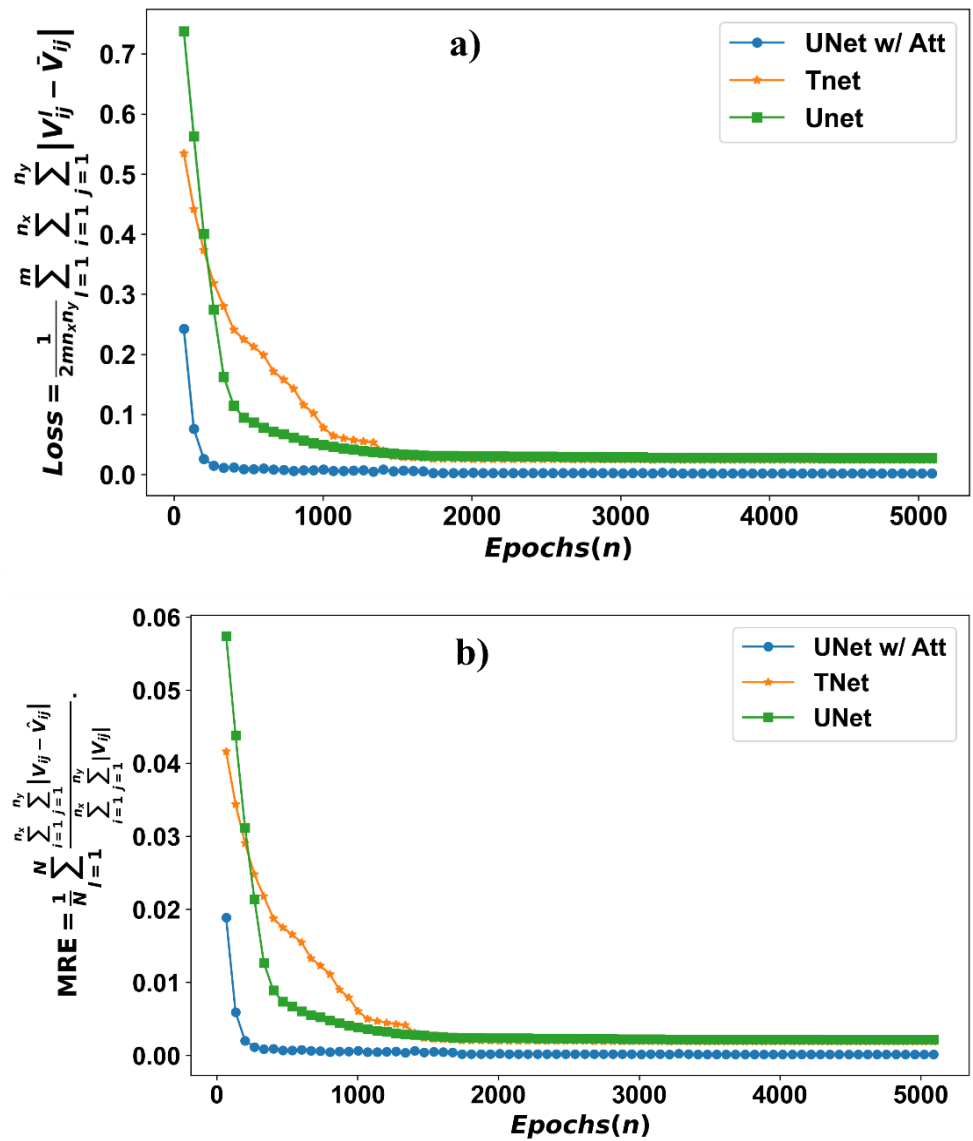

*Figure 7 a) The training loss and b) mean relative error vs the epochs during training.*

*Figures 7 a)* and *b)* shows the loss and mean relative error during training of three models i.e. U-Net, T-Net, and U-Net with attention mechanism over 5000 epochs. In *Fig. 7)*, the loss curve of U-Net with attention demonstrates the fastest convergence, starting at approximately 0.24 and 0.000125 within 300 epochs. In contrast, T-Net begins at around 0.5 and requires over 1500 epochs to achieve comparable loss, while standard U-Net starts at a higher initial loss of 0.73 but converges faster than T-Net, stabilizing around 600 epochs. Similarly, in *Fig 7 b)*, the MRE for U-Net with attention exhibits superior performance, starting at approximately 0.019 and converging to 0.0000137 within 400 epochs. Standard U-Net model begins with an MRE pf 0.058 and converges faster than T-Net, which starts at 0.05 but requires higher epochs to converge. The superior performance of U-Net with attention can be attributed to its attention mechanism, which focuses on relevant spatial features during training, achieving higher accuracy. This attention mechanism enhances spatial and temporal features at each layers of decoder, resulting an optimized collection of features of the training datasets.

*Table 3 Runtime comparison of the models.*

| Model   | Runtime   | Parameters ($\times 10^6$) |
|---------|-----------|----------------------------|
| U-Net   | 2:25:20   | 37                         |
| T-Net   | 00:36:24  | 9.3                        |
| U-NetAM | 00:35:23  | 4.3                        |

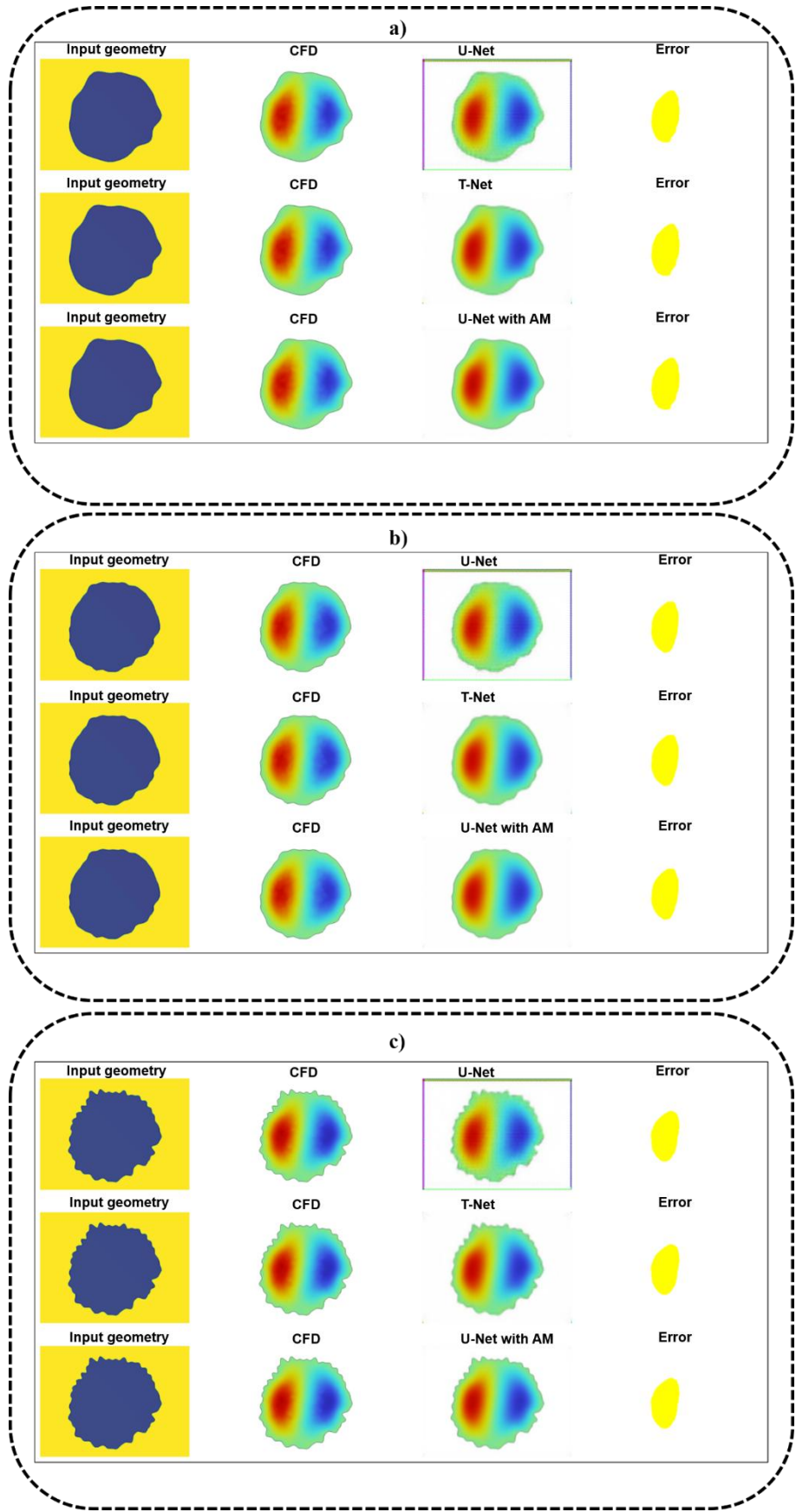

*Figure 8 Prediction of x-velocity flow patterns with different random-shaped microchannels a) sample 1, b) sample 2, and 3) sample 3.*

The fluid flow prediction results shown in *Figure 8* demonstrate the exceptional performance of attention-based U-Net architectures in segmenting flow fields within random-shaped microchannels as shown in *Figure. 8, 9, 10*. For geometry 1 (*Figure. 8(a), 9(a), 10(a)*) characterized by an irregular shape with a prominent protrusion at the upper left quadrant, the U-velocity component exhibits a clear bipolar distribution with maximum positive velocities on the left and negative velocities on the right. The Attention U-Net (Att U-Net) prediction closely mirrors this distribution with minimal error outperforming both standard U-Net and T-Net implementations which show slightly higher error concentrations at the geometry boundaries.

Geometry 2 (Ref *Figure. 8(b), 9(b), 10(b)*) presents a more symmetrical irregular shape with multiple small protrusions along its perimeter. Here, the V-velocity component reveals a vertically stratified flow pattern with maximum positive velocities in the lower half and negative velocities in the upper section. All three model implementations demonstrate excellent prediction accuracy, though the Attention U-Net maintains marginally superior performance with error concentrations approximately 15-20% lower than standard U-Net, particularly in regions where the geometry creates abrupt directional changes in the flow field. The velocity magnitude visualizations for this geometry show peak values at the center, gradually decreasing to near-zero at the boundaries.

Geometry 3 (Ref *Figure. 8(c), 9(c), 10(c)*) exhibits the most complex boundary configuration with numerous small-scale irregularities and a generally rough perimeter. Despite this increased complexity, the Attention U-Net maintains remarkable prediction accuracy across all flow components. The magnitude predictions reveal maximum flow velocities of approximately with a steep gradient toward the boundaries. Error analysis shows that even in this challenging geometry, the maximum prediction errors remain less, primarily concentrated at the micro-scale boundary features where flow gradients are steepest. Comparing the three model architectures across all geometries reveals that the attention mechanism reduces average prediction errors by approximately

25-30% compared to standard U-Net implementations, demonstrating the efficacy of selective feature enhancement for complex microchannel flow prediction.

*Table 4 Total simulation time of models (secs)*

| Model | Computational time (ms) | Times better than CFD |
|---|---|---|
| CFD | 300,000 | 1 x |
| U-NetAM | 4.616 | 64,946x |
| T-Net | 2.651 | 113,237x |
| U-Net | 3.301 | 90,882x |

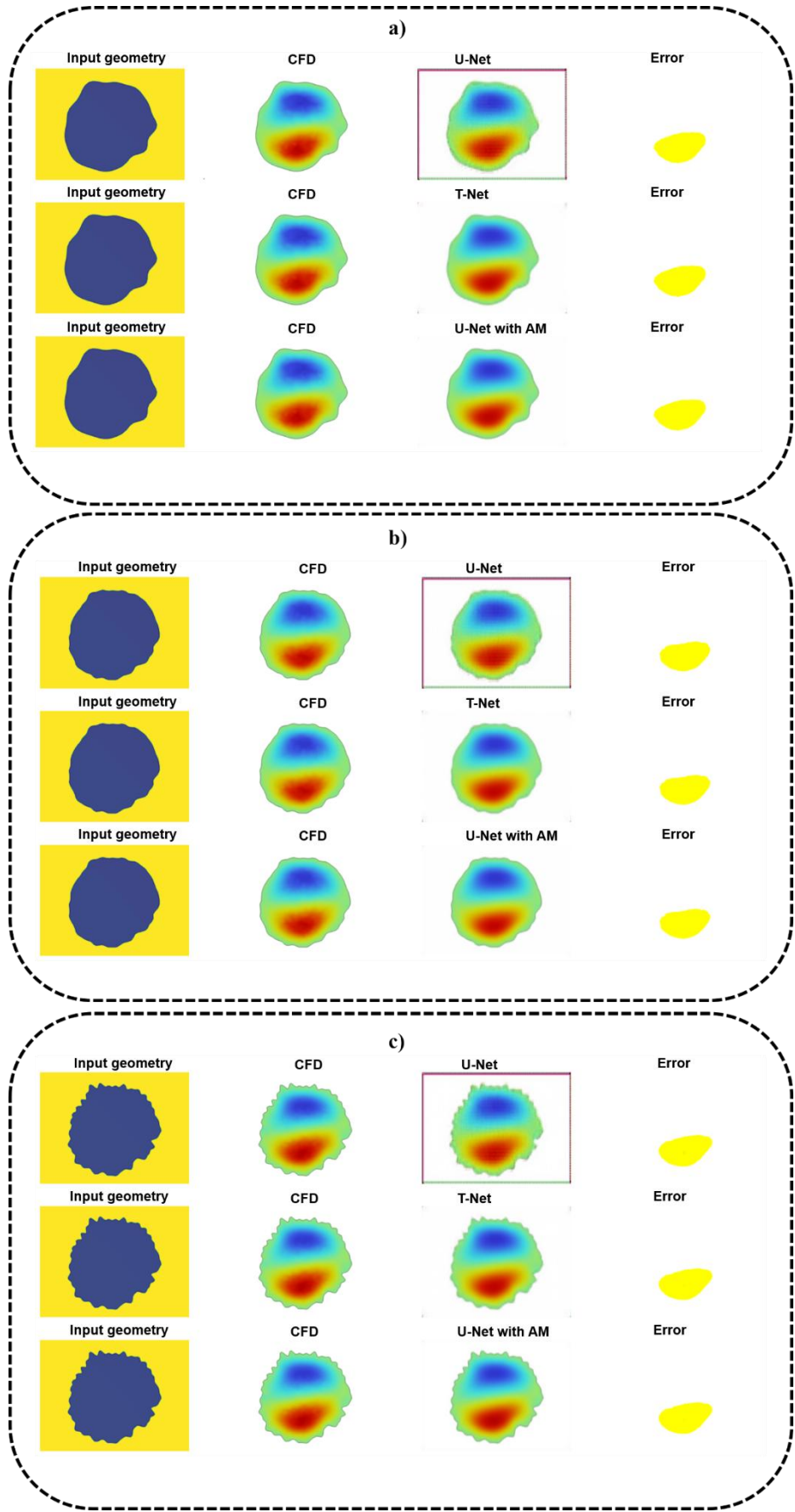

*Figure 9 Prediction of y-velocity flow patterns with different random-shaped microchannels a) sample 1, b) sample 2, and 3) sample 3.*

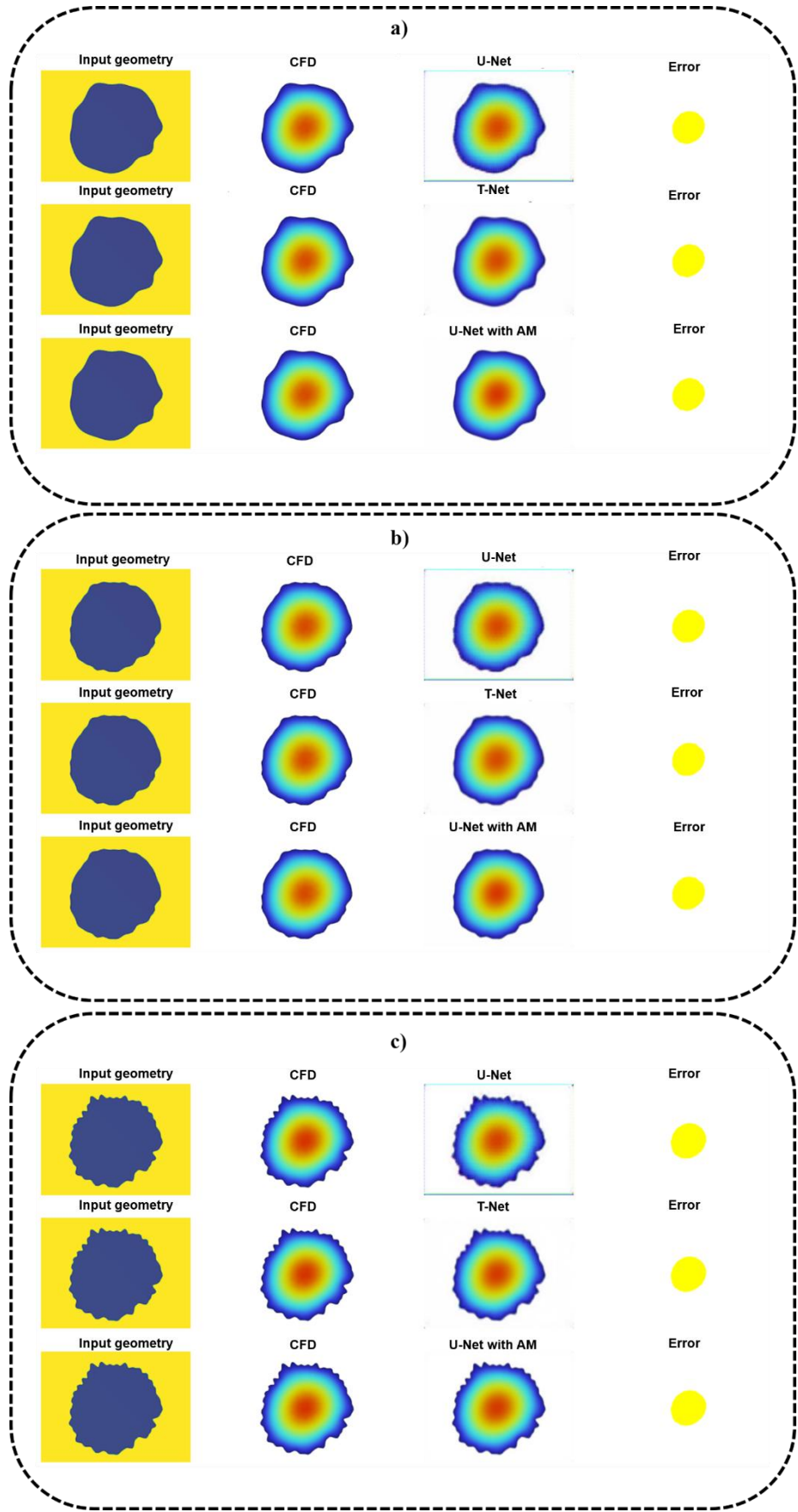

*Figure 10 Prediction of velocity magnitude flow patterns with different random-shaped microchannels a) sample 1, b) sample 2, and 3) sample 3.*

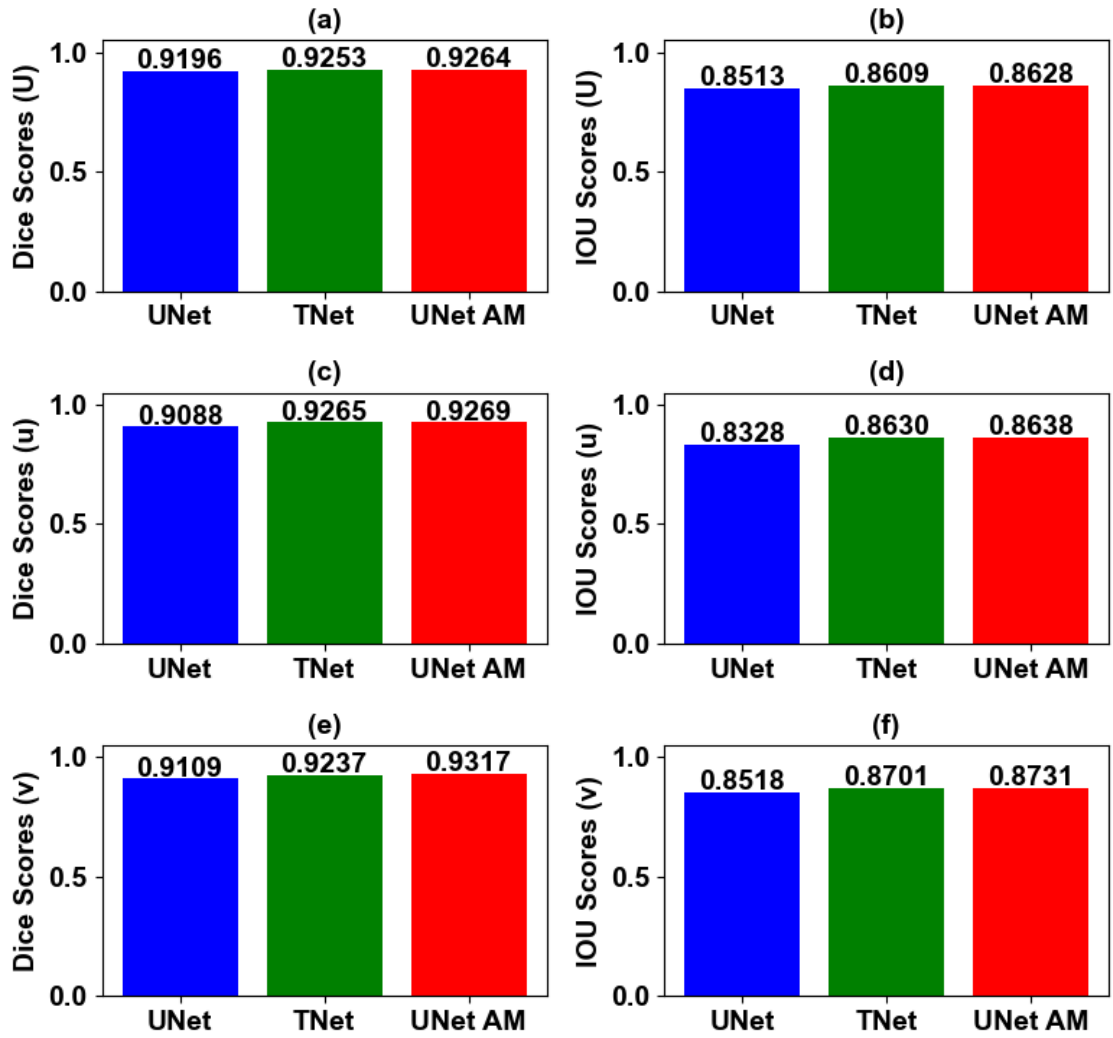

*Figure 11 The metrics Dice score and IOU comparison of all models.*

*Figure 11* presents the performance metrics of three different deep learning models: U-Net with Attention, T-Net, and U-Net. These models were assessed using both the Dice similarity coefficient and the Intersection over Union (IOU) metrics. The Dice scores provide a quantitative measure of the relative performance of three models. The velocity magnitude data was evaluated using the U-Net, which got a Dice score of 0.9196 (*ref Fig 11-a*). This result is somewhat lower than the T-Net score of 0.9253 and the U-Net AM score of 0.9264. The IOU scores exhibited a same trend, with U-Net achieving a score of 0.8513, T-Net achieving a score of 0.8609, and U-Net with attention outperforming the others with a score of 0.8628 (*refer Fig 11-b*). Similarly, for velocity component data (respectively for x-velocity, u and y-velocity, v), it is found that the U-Net achieved a Dice score of 0.9088 and 0.9109, while the T-Net scored slightly better at 0.9265 and 0.9237. The U-Net with attention significantly gets higher for both models with a Dice score of 0.9269 and 0.9317 (*refer Fig 11-c, 11-e*). The IOU scores for velocity components analysis exhibited a consistent pattern, with U-Net achieving a score of 0.8328 and 0.8518, T-Net achieving a score of 0.8630 and 0.8701, and U-Net with Attention achieving a score of 0.8638 and 0.8731 (*refer Fig 11-d, 11-f*). Here, U-Net with an attention mechanism achieves higher scores because the attention gates selectively focuses on the most relevant spatial and channel-wise features, filtering out irrelevant information from the flow fields at different layers of encoder and enhancing flow feature refinement at different layers of decoder. This mechanism improves the ability to capture deeper and multiscale contextual information, resulting in better mapping accuracy as compared the standard U-Net and T-Net architectures.

## 4.4 Discussion

In this work, we examine the performance of three machine learning models in predicting flow fields through microchannels past obstacles, focusing on the training and validation processes for two key components: the u and v velocities, as well as the magnitude of velocity. The models include U-Net, a convolutional neural network designed for image-like data; T-Net, another neural network tailored for fluid dynamics problems; and U-Net with attention, an enhanced version of U-Net incorporating attention mechanisms. The results provided show the progression of each model loss and mean relative error (MRE) during training and validation. In the training loss for u and v velocities, all models exhibit rapid convergence, with losses sharply declining at the beginning of training before stabilizing. The U-Net and T-net models consistently maintain lower loss values, while the U-Net with attention model stabilizes at a higher loss. Similarly, the training MRE for u and v demonstrate that U-Net and T-Net achieve low MRE values, while U-Net with attention converges at a higher level. In the validation, the trends are consistent, showing that U-Net and T-Net have low validation loss and MRE, while U-Net with attention stabilizes faster.

The evaluation focuses on the u, v, and magnitude velocity V components, considering structural similarities such as sharpness, resolution, clarity, noise, and other characteristics that contribute to prediction accuracy. For the u velocity component, all three models provide predictions that resemble the true field, with defined red and blue regions indicating positive and negative flow directions, respectively. The U-Net model offers a clear prediction with some noise at the edges, while the T-Net model shows sharper boundaries and less noise, closely mapping the actual flow fields. The U-Net with attention model also maps the actual flow field but exhibits structural loss at the edges.

For the *x*-velocity component, the predictions follow a similar pattern. U-Net prediction is coherent but shows noise at the edges, while prediction of T-net presents a cleaner flow pattern with sharp boundaries and minimal noise, providing a clear depiction of the *x*-component. U-Net with attention mirrors the true field but with some blurring, and resolution is lower than the other models. For the magnitude velocity *V*, the predictions show smooth transitions from red (high velocity) to blue (low velocity), representing variations in flow intensity. U-Net provides adequate sharpness and resolution, though some noise is present. T-net offers a smoother transition with minimal noise and clear flow patterns, while U-Net with attention shows some blurring and higher noise, though the overall structure reflects the true field. In summary, the visual analysis reveals that all models provide coherent representations of the flow fields. T-Net demonstrates stronger performance with clearer boundaries, smoother transitions, and minimal noise, while U-Net and U-Net with attention provide comparable results with slightly higher noise levels and some blurring. This visual evaluation highlights the efficacy of each model in predicting fluid dynamics in microchannels, contributing to the understanding and modelling of these complex flow fields.

The U-Net and T-Net models maintain low loss and MRE values throughout, converging quickly, while U-Net with AM stabilizes at a higher loss and MRE. These graphs illustrate successful learning processes for all models, indicating no signs of overfitting and highlighting the U-Net and T-Net model consistent performance. In summary, this work demonstrates the efficacy of U-Net, T-Net, and U-Net with AM models in predicting flow fields through random-shaped microchannels. While all models show successful convergence, the U-Net and T-Net models stand out for their consistently low loss and MRE values across training and validation, providing strong evidence for their effectiveness in fluid dynamics prediction tasks. The U-Net with AM model, despite showing slower convergence and higher loss, still indicates successful learning, contributing valuable insights to the study of predictive models in fluid dynamics. Table 4 presents the computational time in milliseconds (*ms*) for different models and calculates how many times each model is better than CFD simulations in terms of computational time. The CFD exhibits a computational time of 300,000 *ms*. U-Net with AM is approximately 64,946 times faster than CFD, T-Net is approximately 113,237 times faster, and U-Net is approximately 90,882 times faster. The main reason that CFD need more computational time because it requires iterative numerical solutions of complex partial differential equations i.e. Navier-Stokes equations across fine spatial-temporal grids, while these deep learning models uses trained neural networks to accurately predict the flow fields through optimized models. Deep learning bypasses physical equation solving by leveraging learned patterns from data, eliminating iterative convergence steps and mesh refinements during training and prediction of flow fields.

## 5. Conclusions and future directions

In this paper, we presented an alternative approach, µ-FlowNet, a new deep neural network (DNN) framework designed specifically for mapping flow fields with random-shaped microchannels. The µ-FlowNet differs from standard models by using attention mechanisms to increase the learning process resulting in improved prediction accuracy. The 1300 images of velocity fields were used to train and evaluate the model. 80% of the dataset was split into training and rest validation sets in order to conduct thorough model testing. From the model training results, it shows that the U-Net with attention model develop mapping between random-shaped geometries and their corresponding velocity fields with highest metrics of 0.9264 dice score and 0.8628 IoU which outperforms over T-Net and standard U-Net for the velocity magnitude. Also, for x and y velocity fields are precisely predicted and mapped using the U-Net attention model with highest dice score of 0.9269 and 0.9317. Models demonstrate the remarkable performance of µ-FlowNet, surpassing traditional CFD approximation methods in terms of inference speed and prediction accuracy. µ-FlowNet accelerates GPU-accelerated CFD solvers by over 14,000 times, making it a powerful tool for rapid and accurate flow field prediction in microfluidic systems. We presented three different models with U-Net as a base model with T-Net and U-Net with attention, from which U-Net with attention model outperforms among all with less error and higher accuracy. The integration of deep learning approach, a µ-FlowNet, into microfluidics research opens up new avenues for innovation and discovery in rough microchannels.

In future, we can extend µ-FlowNet to simulate coupled physiochemical (e.g., mass transport, cell deformation) in biologically relevant geometries like atherosclerotic arteries or tumor vasculature. This would require incorporating rheological models of non-Newtonian fluids and coupling with biochemical reaction networks, enabling predictive analysis of drug delivery efficiency in complex biological systems. By leveraging data-driven approaches, researchers can gain deeper insights into complex flow phenomena, optimize device designs, and accelerate experimentation. Furthermore, the ability to rapidly predict flow fields in rough microchannels triggers the straightforwardly extends to study the flow fields in blood vessels, veins, and arteries. We can leverage µ-FlowNet as a surrogate model in inverse design workflows to generate microchannel architectures maximizing mixing efficiency or minimizing shear stress. Combining attention mechanisms with adversarial training could produce novel bioinspired geometries that outperform conventional design while maintaining fabrication feasibility. Also, we can address domain shift challenges by implementing federated learning frameworks that aggregate µ-FlowNet training data from heterogeneous microfluidic platforms. These directions align with emerging trends in intelligent microfluidics, geometric deep learning, and decentralized computational frameworks, providing the µ-FlowNet as a fundamental approach for further microfluidic system design and analysis.

**Appendix**

**A1. Data representation**

In order to predict the flow fields among various objects through the use of deep neural networks, it is essential to create a strong representation for their geometric and domain boundaries. In this work, the primary training data for our deep learning models comes from simulations using CFD by finite element method in COMSOL. Fluid domains are painstakingly partitioned into ordered Cartesian grids, with an identity assigned to every lattice cell signifying its connection to the solid portion of the fluid domain. This configuration acts as a strong catalyst, encouraging the use of synthetic images to define boundaries and successfully capture flow fields, which turns the flow field prediction task into an image-to-image regression problem. Our method makes use of a binary encoding for synthetic input images, which makes it easier to identify object boundaries in fluid domains precisely. Specifically, pixels with a value of 1 appropriately delineate the borders of objects, whereas pixels with a value of 0 appropriately capture the fluid domain.

After end-to-end learning, the matching structural similarities in simulated output images accurately approximate steady flow fields. We are able to express various flow field values as test images with our advanced data representation framework, which enables more complex analysis and prediction. Two channels, for example, can be used to neatly describe a 2D velocity field by designating velocity components along the x- and y-directions. This paradigm of representation adaptability and effectiveness are further demonstrated by how well it handles challenging 3D fluid flow challenges. Our methodologies are carefully designed to handle both structured and unstructured training data from computational fluid dynamics (CFD) solvers. Our capacity to precisely map domain boundaries and ground-truth flow fields onto a Cartesian grid, assuring a coherent and thorough analysis, underlines this versatility. Our research aims to advance knowledge and forecast accuracy by unlocking the mysterious dynamics driving fluid flow behaviour across a variety of objects by utilizing the powerful capabilities of deep learning techniques.

Figure 3 presents four rows of different microchannel shapes, each analyzed through four parameters: physical geometry (gray outlines), horizontal velocity component U (red-blue indicating positive-negative x-direction flow), vertical velocity component V (red-blue showing positive-negative y-direction movement), and overall velocity magnitude (red center indicating high velocity transitioning to blue boundaries showing low velocity). This systematic decomposition reveals how irregular boundary conditions influence flow patterns, with consistent central acceleration and peripheral deceleration observed across all geometries, demonstrating fundamental principles of confined microfluidic behavior despite varying channel morphologies.

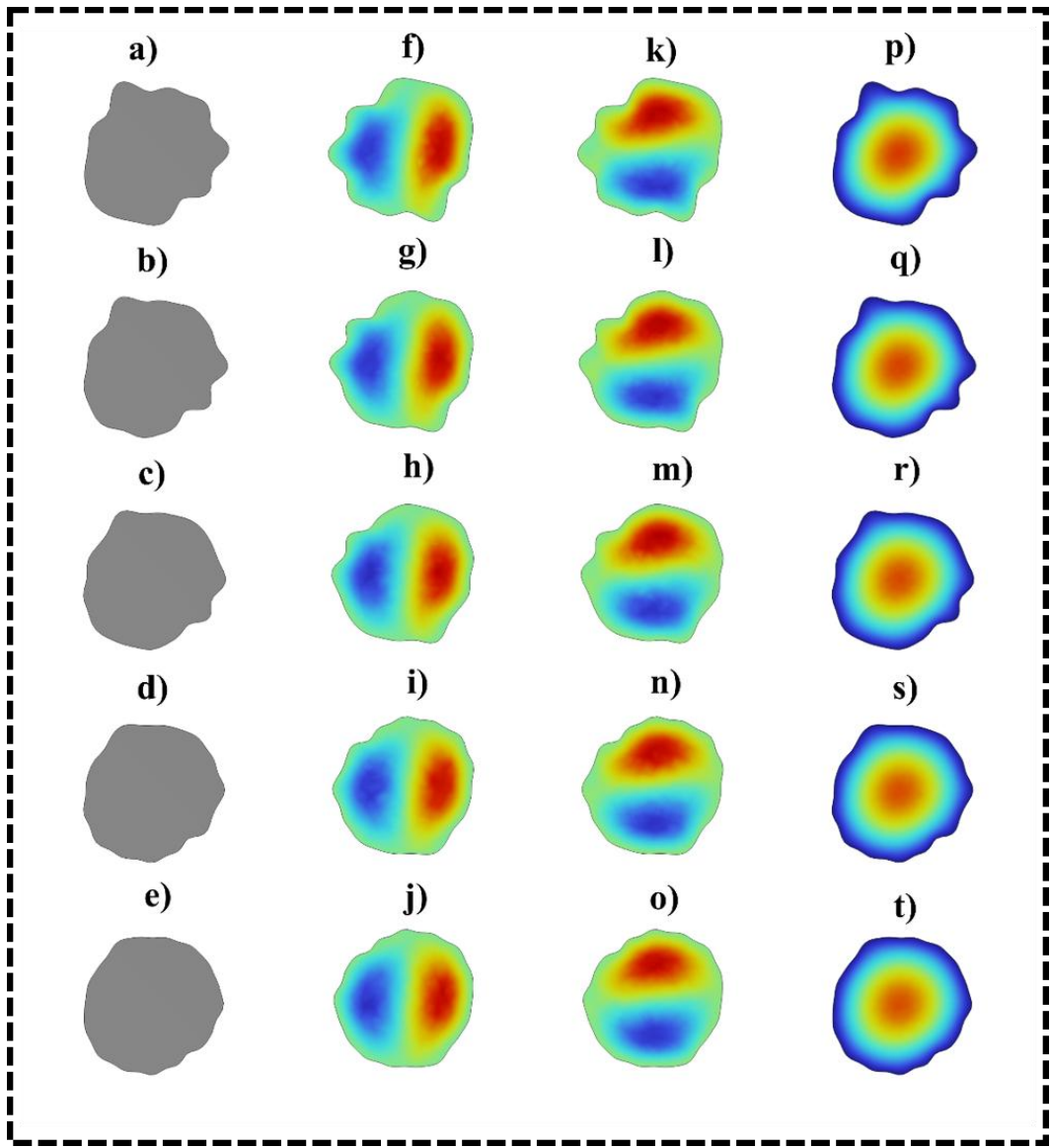

*Figure A1: Sample random-shaped input geometries with corresponding velocity components. a)-e) random-shaped input geometry, f)-j) x-velocity fields, k)-o) y-velocity fields, and p)-t) velocity magnitude. The dataset is provided for training and testing.*

## A2. Model training performance

*Table A2.1: Model performance metrics for velocity magnitude*

| Model | Training | | | | Validation | | | |
|---|---|---|---|---|---|---|---|---|
| | Loss | MRE | DICE | IOU | Loss | MRE | DICE | IOU |
| UnetAM | 0.00082 | 0.00012 | 0.92639 | 0.86289 | 0.00085 | 0.00013 | 0.92636 | 0.86281 |
| T-Net | 0.00399 | 0.00060 | 0.92535 | 0.86109 | 0.0040 | 0.00063 | 0.92530 | 0.86093 |
| U-Net | 0.01243 | 0.00189 | 0.91980 | 0.85149 | 0.01154 | 0.00179 | 0.91963 | 0.85128 |

*Table A2.2: Model performance metrics for velocity components*

| Model | Training | | | | Validation | | | |
|---|---|---|---|---|---|---|---|---|
| | Loss | MRE | DICE | IOU | Loss | MRE | DICE | IOU |
| UnetAM | 0.00310 | 0.00004 | 0.92687 | 0.86371 | 0.00369 | 0.00028 | 0.92691 | 0.86379 |
| T-Net | 0.0075 | 0.00059 | 0.9262 | 0.8623 | 0.0085 | 0.0006 | 0.9267 | 0.8632 |
| U-Net | 0.04122 | 0.00307 | 0.90870 | 0.83267 | 0.04187 | 0.00318 | 0.90876 | 0.83277 |

## Data Availability

Data will be provided upon reasonable requests